\documentclass[final,1p,times]{elsarticle} 
\usepackage{graphicx} 
\usepackage{amssymb} 
\usepackage{amsthm} 
\usepackage{lineno} 

\usepackage{amsmath}

\newcommand{\qs}{Q_\mathrm{s}}
\newcommand{\lqcd}{\Lambda_{_{\rm QCD}}}
\newcommand{\as}{{\alpha_\mathrm{s}}}

\newcommand{\mev}{\textrm{ MeV}}
\newcommand{\gev}{\textrm{ GeV}}

\newcommand{\ra}{R_{_{\rm A}}}
\newcommand{\rp}{R_{\rm p}}

\newcommand{\ud}{\mathrm{d}}
\newcommand{\fig}{fig.~}

\newcommand{\se}{sec.~}

\journal{Nuclear Physics A} 
\begin{document} 

\begin{frontmatter} 


\title{Understanding saturation and AA collisions with an eA collider}

\author{T. Lappi}

\address{Department of Physics
P.O. Box 35, 40014 University of Jyv\"askyl\"a, Finland and \\
Institut de Physique Th\'eorique, CEA/DSM/Saclay,
91191 Gif-sur-Yvette Cedex, France
}

\begin{abstract} 
The initial conditions in high energy nucleus-nucleus collisions are determined
by the small momentum fraction part of the nuclear wavefunction.
This is the regime of gluon saturation and the most direct way to experimentally
study it would be
deep inelastic scattering at a high energy electron ion collider (EIC).
This talk discusses some of the connections between physics at the EIC
and the initial stage of relativistic heavy ion collisions.
We argue that measurements at an EIC will provide detailed high-precision
information about the parameters for the initial conditions, transverse geometry
and longitudinal correlations that will be crucial in understanding
the initial stage of a heavy ion collision.
\end{abstract} 

\end{frontmatter} 



\section{Introduction}\label{sec:intro}
Hadrons and nuclei consist of partons, and the relevant degrees of 
freedom for describing their collisions at high energy are the quark and gluon 
fields. The most convenient kinematical variables to describe these
degrees of freedom are the momentum transfer $Q^2$ (interpreted as a
resolution scale in the transverse direction) and $x$, the fraction
of longitudinal momentum carried by the parton in a frame where the
hadron momentum is large. In a collision of two nuclei or protons it is 
usually not possible to experimentally determine  the precise values
of $x$ and $Q^2$ that were involved in the production of a final state particle;
one measures only
convolutions of the properties of the two wavefunctions. In deep inelastic
scattering (DIS), by contrast, the outgoing electron is measured and it
is therefore possible to know exactly the values of $x$ and $Q^2$
probed from the measured momenta. This is the feature that makes DIS the ideal
way to obtain precision information about the QCD wavefunction.
 
With the planning process underway to build an Electron-Ion-Collider 
(EIC)~\cite{Deshpande:2005wd} or, on a somewhat longer timescale,
a Large Hadron-electron Collider (LHeC)~\cite{Dainton:2006wd} the subject is a
topical one.
This talk is, however, not intended to be a review of the experimental 
program at the EIC or the LHeC. We will instead 
concentrate on a few particular aspects in which nuclear DIS (nDIS) experiments
can and already have been useful in understanding the initial stages
of a heavy ion collision and therefore crucial for experimentally studying the
properties of the Quark-Gluon Plasma.

\section{The nonlinear high energy regime of QCD}\label{sec:sat}

\begin{figure}
\begin{center}
\resizebox{0.99\textwidth}{!}{
\includegraphics[height=5cm,clip=true]{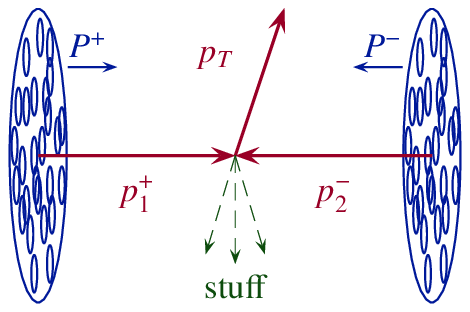}
\rule{0.2cm}{0pt}
\includegraphics[height=5cm,clip=true]{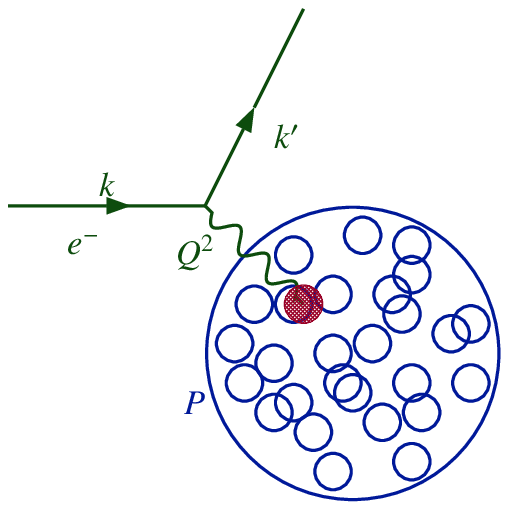}
\rule{0.2cm}{0pt}
\includegraphics[height=5cm,clip=true]{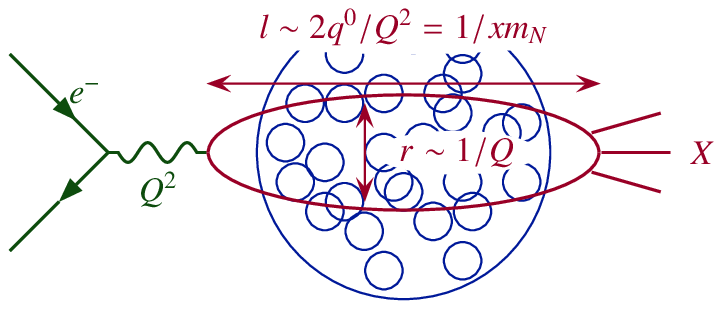}
}
\end{center}
\caption{
Left: Kinematics in hadronic or nuclear collisions. The momenta of the incoming partons
$p_1^+$ and $p_2^-$ cannot be reconstructed based on the measured $p_T$ and the beam 
energies $P^+$ and $P^-$ alone because of the recoil of the underlying system.
Center: The kinematics in DIS: the incoming momenta $P$ and $k$ and the outgoing electron 
$k'$, are measured, and the kinematics of the struck parton can be reconstructed from these
values. Right: DIS at small $x$ in the dipole frame. The virtual photon fluctuates into a
quark-antiquark dipole with lifetime $\sim 1/x$ and transverse size $r \sim 1/Q$.
} \label{fig:pstricks}
\end{figure}

In collisions of protons and nuclei the typical values of $x$ that are probed in the 
wavefunction are $x \sim p_\perp/\sqrt{s}$. Let us consider, in the center-of-mass
frame of the collision, the nucleus moving in the $+z$ direction.
In light cone variables $p^{\pm} = (p^0  \pm p^3)/\sqrt{2}$ and
$x^{\pm} = (t  \pm z)/\sqrt{2}$  we consider $p^+$ as
the longitudinal momentum, $p^-$ as the light cone energy, $x^+$ the light cone time
and $x^-$ the longitudinal coordinate. Note that the variables $x^\pm$ are conjugate
to $p^\mp$.
 The parton 
with momentum fraction $x$ will have longitudinal
momentum $x p^+ \sim x \sqrt{s}/A$ and will 
thus probe the other, leftmoving, nucleus at a length scale $\Delta x^-  \sim A/ (x \sqrt{s})$.
The  longitudinal size of the leftmover is Lorentz-contracted from $\ra \sim A^{-1/3} \rp$
to $\sim A^{-1/3} \rp ( A m_N/\sqrt{s})$. We see that if $x \ll A^{-1/3} \rp  m_N$, the
partons in the rightmoving nucleus will not be able to resolve the individual nucleons 
of the leftmoving one. The whole nucleus must therefore be treated as one coherent
target, not as a collection of independent nucleons. The observation that the
large $x$ localized, valence-like, degrees of freedom are not resolved in the collision, but only
the smaller $x$ partons that they radiate, naturally leads to the idea of treating the two
separately in an effective field theory approach. This effective field theory is 
known as the Color Glass Condensate (CGC). 
We refer the reader to the reviews~\cite{Iancu:2003xm,Weigert:2005us}
for further details, but it suffices here to emphasize the following.
The CGC describes a high energy hadron in terms of a classical strong color field 
(the small $x$ gluons) radiated by an effective colour current (the large $x$ degrees of
freedom). The classical color charges are stochastic random variables
with a probability distribution $W_x[\rho]$.
 Both the initial field configuration in a heavy ion collision
and observables in DIS at small $x$ can be computed in terms of these 
same classical gluon fields.
The color charge distribution depends on nonperturbative input and cannot 
completely be computed from first principles. Its dependence on the energy 
scale (rapidity) that separates the large and small $x$ degrees of freedom can, 
however, be computed and expressed in terms of a renormalization group equation.
The distribution of color charges is a universal object, it can be measured 
in one process (ideally DIS) and then used as an independent input to 
make prediction for another one (say, the initial field configurations
in a heavy ion collision). In this sense the situation is analogous to 
collinear factorized perturbation; there is a universal, nonperturbative
distribution (color charge distribution or parton distribution function),
a separation scale (rapidity or virtuality) and a renormalization group equation
derived from first principles that describes the dependence on this
separation scale.

Gluon saturation appears as a very different phenomenon in different Lorentz
frames. In the infinite momentum frame, where the parton model is defined,
saturation arises from nonlinear interactions between gluons. A more convenient 
description of DIS at small $x$ is obtained in the dipole frame (roughly the 
target rest frame). In this frame the process can be viewed as a virtual photon
with four-momentum $q$ and virtuality $q^2 = -Q^2$ 
splitting into a quark-antiquark dipole that then interacts 
with the target (see \fig\ref{fig:pstricks}).
At high energy (or small $x$) this fluctuation 
has a lifetime $\sim  1/(xm_N)$ which, for the values 
of $x$ that we are considering, is much larger than the size 
of the nucleus. The dipole therefore does not resolve individual nucleons,
but interacts coherently  with the nucleus as a whole.

The transverse size of the dipole $r$ is related, by the wavefunction of the
virtual photon splitting into a fermion pair; to the momentum transfer, 
$r \sim 1/Q$.
The interaction between the dipole and the target is described by a 
scattering amplitude $\mathcal{N}(x,r,\boldsymbol{b}_\perp)$ or, integrated over the impact
parameter $\boldsymbol{b}_\perp$, the dipole cross section
$\sigma(x,r) = 2 \int \ud^2 \boldsymbol{b}_\perp\mathcal{N}(x,r,\boldsymbol{b}_\perp)$.
In the limit of small dipole sizes the scattering amplitude should vanish, because
a dipole of size $r=0$ is s colorless object. For small $r$ the dipole
scattering amplitude behaves as $\sim r^2 xG(x,Q^2 \sim 1/r^2),$
where $xG(x,Q^2 \sim 1/r^2)$ is the conventional integrated gluon distribution.
For large $Q^2$ this behavior dominates, and one recovers back the DGLAP description
applicable in the dilute regime. The scattering amplitude is, however, bound by unitarity:
$|\mathcal{N}(x,r,\boldsymbol{b}_\perp)| \leq 1$. The growth as a function of $r$ 
cannot, therefore, continue indefinitely. It must be modified for 
large $r$, i.e.  for $Q^2$ smaller than some characteristic scale 
$\qs^2 \sim xG(x,Q^2 \sim \qs^2)$;
the \emph{saturation scale}. In the infinite momentum frame this characteristic 
scale corresponds to the typical transverse momentum of the gluons in the wavefunction.
Furthermore, it is observed experimentally and understood theoretically in terms
 of an exponentially growing cascade of bremsstrahlung gluons that the gluon distribution
rises strongly at small $x$ as $xG(x,Q^2) \sim x^{-\lambda}$. This leads to the conclusion 
that the saturation scale must rise as a function of energy, typically as 
$\qs^2 \sim x^{-\lambda}$. If the collision energy is too small, the saturation scale
is $\sim \lqcd$ and weak coupling methods can only be applied to rare high $Q^2$ phenomena,
not the bulk dominated by $\qs$. For large enough energies, however, $\qs \gg \lqcd$ 
and weak coupling methods can be used.

In the infinite momentum frame the same phenomenon of gluon saturation looks completely 
different. There it is easy to understand the growth of the gluon distribution with energy;
with more phase space available for radiation the gluons tend to split and their number grows 
exponentially with rapidity, i.e. as a negative power of $x$. When the phase space density
of the gluons becomes of order $1/\as$, the nonlinear interactions among them become important
and they start to recombine, which slows the growth of the gluon distribution. The curious 
thing about these two views of the same phenomenon is that in what in the infinite momentum 
frame looks like the result of nonlinear interactions and is more complicated to quantify 
is in the
dipole frame a simple and precise statement based on the unitarity of the $S$-matrix. In many 
treatments of high energy evolution and saturation in the dipole frame the effect of
increasing the energy is treated as a boost given the dipole (Lorentz-invariance guarantees
that one can choose to boost either the dipole or the target as convenience dictates,
the physical result must be the same), which makes the dynamics appear as splitting and merging 
of dipoles in the probe, instead of gluons in the target as in the infinite momentum frame.
From the discussion in the dipole frame it should be clear that the relevant question is not 
whether parton saturation exists;
the saturation in the gluon distribution at some energy-dependent transverse scale
is required by unitarity. At asymptotically high energies this scale is bound to be large
enough compared to $\lqcd$ for a weak coupling description of the process to be applicable.
The relevant question is instead what the value of $\qs(x)$ is, and whether the momenta and
energies of the process one is studying are close enough to $\qs$ that saturation
has to be taken  into account. As we shall argue in the following, there is strong evidence 
that this is the case for bulk particle production and forward jet production at RHIC, and
for most of the properties of the initial state in heavy ion collisions at the LHC.

\section{Parameters of the initial condition for AA}\label{sec:numbers}

\begin{figure}
\resizebox{\textwidth}{!}{
\includegraphics[height=5cm,clip=true]{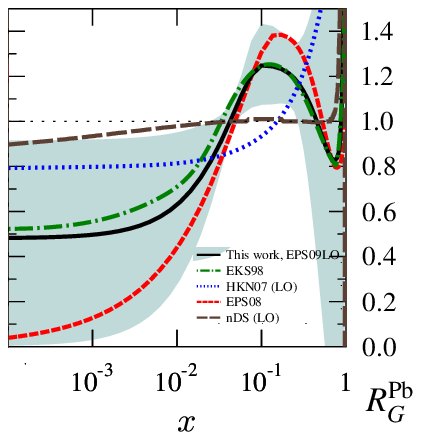}
\rule{1cm}{0pt}
\includegraphics[height=5cm,clip=true]{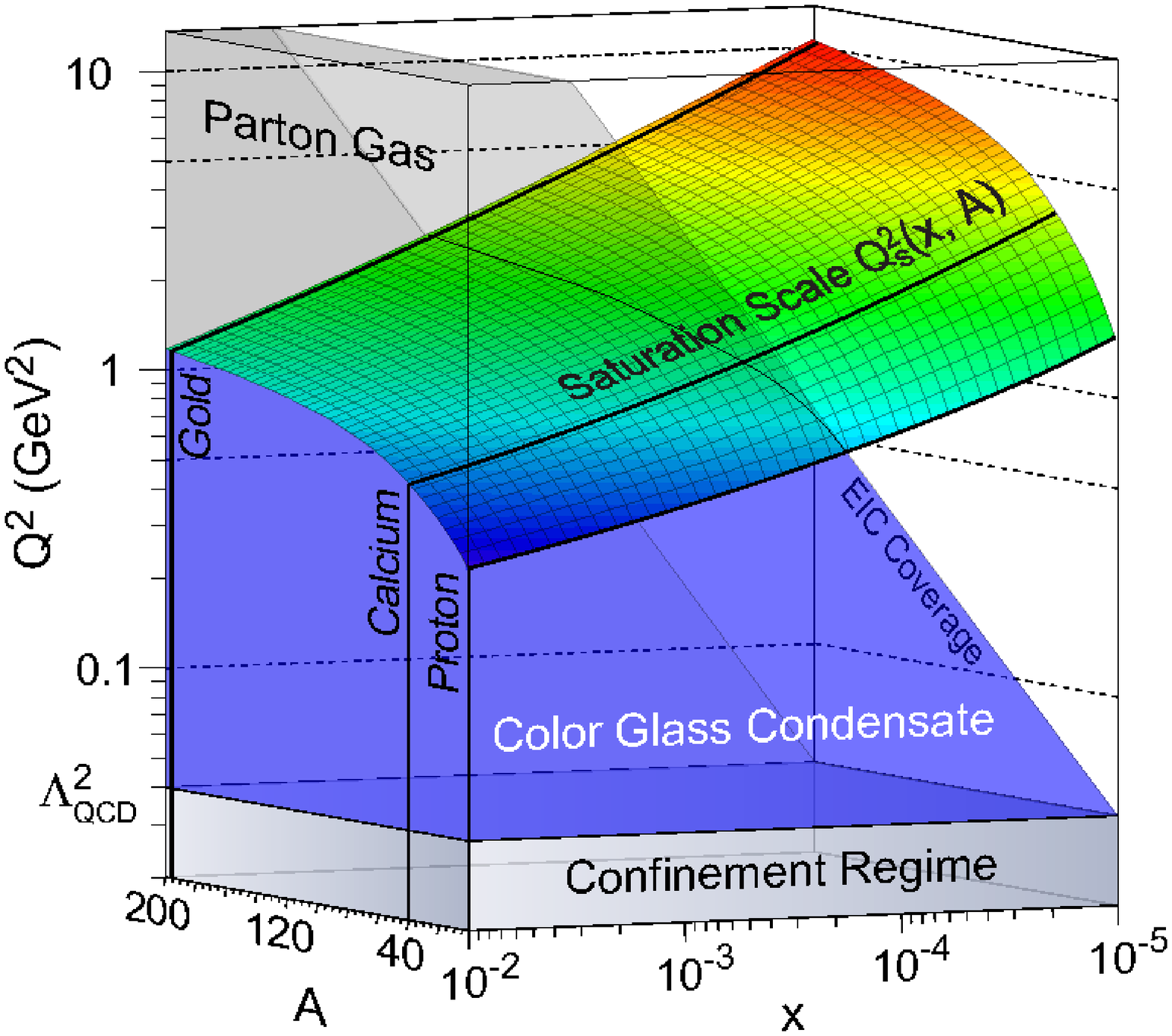}
}
\caption{Left: Different parametrizations of the nuclear modification to the gluon distribution
at $Q^2=1.69\gev$; plot from Ref.~\cite{Eskola:2009uj}. The differences between the
parametrizations at larger $Q^2$ are not quite as drastic, though.
Right: The dependence of the saturation scale $\qs$ on $x$ and the 
mass number $A$.} \label{fig:eps}
\end{figure}

An important measure of the properties of the quark gluon plasma
are the properties of hard particles or jets propagating though the medium. 
In order to isolate the effects of the medium one needs to be able to calculate
the production rates of hard partons via pQCD methods, which requires
knowing the nuclear parton distribution functions. Because there is so little 
data on nDIS, these, especially the gluon distribution are poorly constrained 
at small $x$; this is demonstrated in \fig\ref{fig:eps}. At RHIC
the $x$ values relevant for jet production are, apart from  forward rapidities,
still quite large, but this will change significantly at LHC energies, where
partons with $x \ll 10^{-2}$ can produce jets with $p_\perp > 10 \gev$. This means
that the jet production is very sensitive to a region in $x$ where
the nuclear gluon distribution is virtually unknown.

\begin{figure}
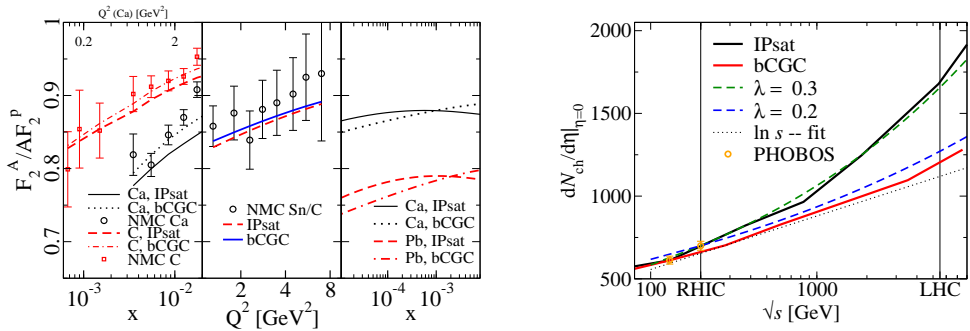

\centerline{\resizebox{0.95\textwidth}{!}{
\includegraphics[height=5cm,clip=true]{shadcombowbcgcsncwide}
\rule{1cm}{0pt}
\includegraphics[height=5cm,clip=true]{enscan1}
}}
\caption{
Left: Comparison of two impact paremeter dependent dipole model parametrizations 
fitted to HERA data~\cite{Kowalski:2006hc} to nDIS data.
Right: Calculations of the charged multiplicity in in heavy ion collisions from 
RHIC to LHC energies from~\cite{Lappi:2008eq}. Shown are 
calculations using the same dipole cross section parametrizations, extrapolations using a constant
$\lambda$ in $\qs^2 \sim x^{-\lambda}$ and a logarithmic fit to RHIC
data.}\label{fig:nmc} \label{fig:enscan}
\end{figure}

Bulk particle production in heavy ion collisions is dominated by the 
saturation scale $\qs$. For a genuinely independent understanding 
of the initial conditions of a heavy ion collision we should be able to 
determine its value independently, without relying on modeling of the
later stages of the evolution. The evolution of the saturation 
scale with the collision energy can be computed perturbatively, but 
its actual value depends on the initial condition, which is a 
nonperturbative input that has to be obtained from experimental data.
This can be done using DIS experiments. 
Dipole model fits to HERA data constrain the saturation scale
in a proton, which can then be combined with basic nuclear geometry
to calculate $\qs$ in a large nucleus~\cite{Kowalski:2003hm} 
(see \fig\ref{fig:eps}).  The existing nDIS data is from
a too small energy range to provide a very stringent constraint
on the nuclear $\qs$, although it has been used in some attempts to 
parametrize the $A$ dependence of $\qs$~\cite{Freund:2002ux,Armesto:2004ud}.
A comparison of two parametrizations fitted to HERA data and extended
to nuclei~\cite{Kowalski:2003hm,Kowalski:2007rw} is shown in \fig\ref{fig:nmc}.
For the central rapidity region at RHIC this estimate yields a value
$\qs \approx 1.2\gev$. This value can then be used to compute
the initial gluon multiplicity using either a numerical solution of
the classical Yang-Mills 
equations~\cite{Krasnitz:1998ns,Krasnitz:2001qu,Lappi:2003bi,Krasnitz:2003jw}
(see~\cite{Lappi:2007ku} for a discussion on relating the numerical
value of $\qs$ between CYM calculations and DIS observables) 
or in a $k_\perp$-factorized perturbative 
approximation~\cite{Kharzeev:2000ph,Kharzeev:2001gp}, with the 
result of somewhere around $1000$ gluons per unit rapidity in the initial
state of a hevy ion collision. The remarkable 
fact about this value is how well it fits in with a picture of 
fast thermalization and subsequent ideal, i.e. entropy-conserving,
hydrodynamical evolution of the system. One could say that this agreement
is even too good for a leading order calculation since it leaves so little
room for higher order contributions and an increase in the entropy from
the thermalization phase.

When trying to extrapolate these ideas to the LHC, the situation is less settled, 
however. Initial stage gluon production is, to a very good accuracy, a one scale problem. 
The number of gluons in the initial stage of heavy ion collisions must then be
$\approx c \qs^2(x) \pi \ra^2/\as,$ where $c$ is a nonperturbatively determined constant.
The dependence of the multiplicity and initial energy density follows from the $x$-dependence of
the saturation scale. Estimates from fits to HERA inclusive
data (e.g.~\cite{Golec-Biernat:1998js,Bartels:2002cj,Iancu:2003ge,Kowalski:2006hc})
vary in the range $\lambda= 0.2 \dots 0.3$. 
Running coupling BK leads to a $\lambda$ varying with $x$, but within the same range for
the energies between RHIC and LHC~\cite{Albacete:2007sm}.
Extrapolated over the wide range of energy separating LHC from RHIC this turns into
a major uncertainty on the prediction for the LHC initial gluon multiplicity,
see \fig\ref{fig:enscan}. Turning this argument around means that the measurement
of the charged hadron multiplicity in AA-collisions at the LHC will provide a relatively
simple and clear constraint on the interpretation of the HERA data. Disentangling
the many effects that influence the energy dependence of $\qs$ would greatly benefit from 
precision measurements at the EIC.

\section{Diffraction and transverse geometry}\label{sec:tr}
\begin{figure}
\centerline{\resizebox{0.9\textwidth}{!}{
\includegraphics[width=0.56\textwidth,clip=true]{f2d3casnauipsatbcgc5gev2xbj0_001}
\hfill
\includegraphics[width=0.4\textwidth,clip=true]{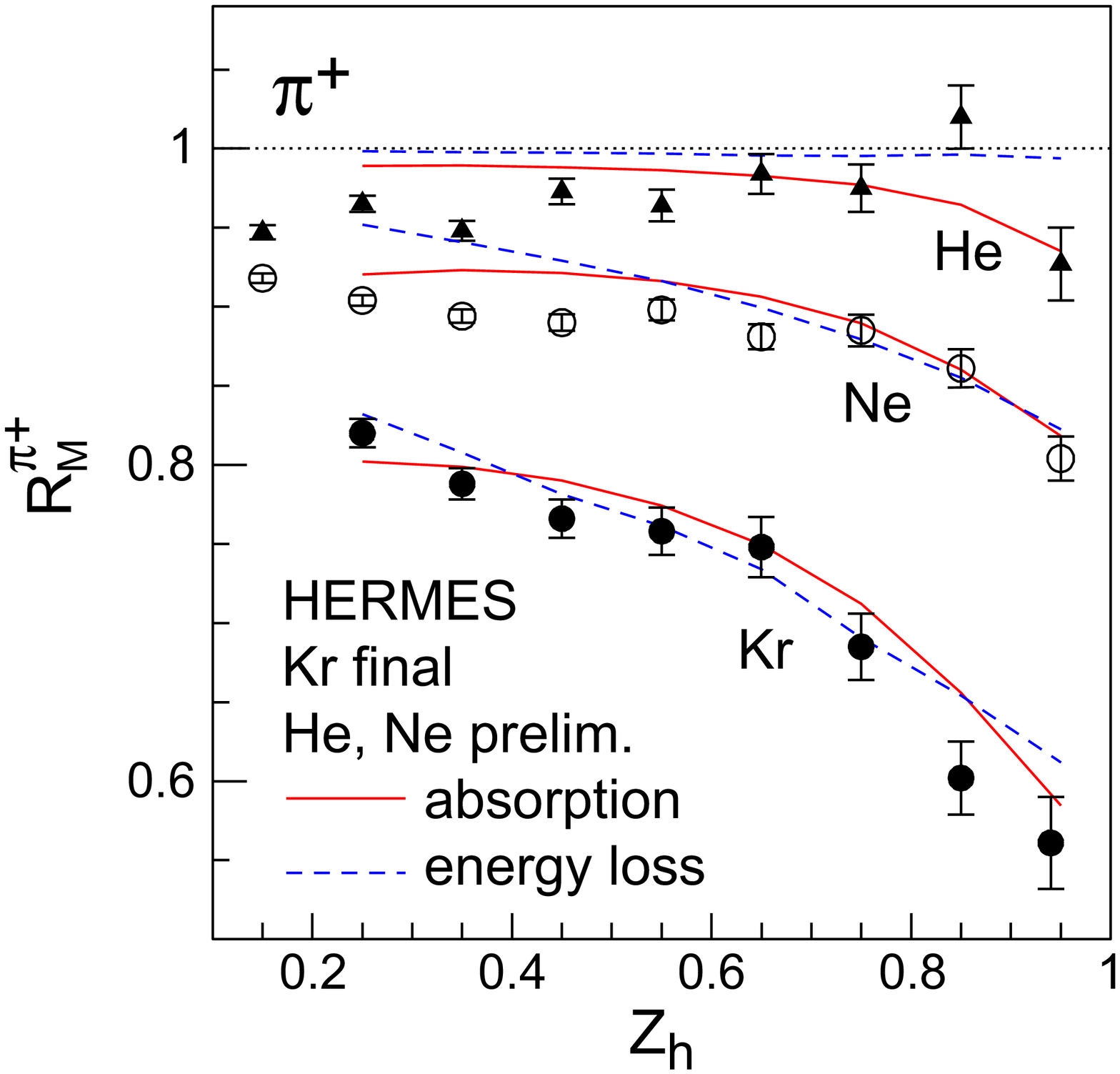}
}}
\caption{
Left: Diffractive structure functions in nuclei compared to the proton, scaled by 
$A$~\cite{Kowalski:2008sa}. The additional kinematic variable introduced here
is $\beta = Q^2/(Q^2-t+M_X^2)$.
Right: Nuclear modification to $\pi^+$ production due to parton
energy loss in cold nuclear matter~\cite{Accardi:2005mm}.}\label{fig:diff}\label{fig:eloss}
\end{figure}

A major discovery at HERA was that a large fraction ($\sim$ 15\%) of high energy DIS 
events are \emph{diffractive}. In this context a diffractive event is defined as 
one where the virtual photon exchanges momentum with the target and 
dissociates into a hadronic system of invariant mass $M^2_X$ (e.g. a vector meson) 
with the target staying 
intact. The experimental signature of these events is that the diffractive system is 
separated from the target by a large rapidity gap with no produced particles, indicating
that no quantum numbers (in particular no color charge) has been exchanged. 
Diffractive DIS has a natural interpretation in the dipole picture, where it corresponds
to elastic scattering between the dipole and the target. In the unitarity 
limit (and when the scattering amplitude is purely imaginary, as it is to a 
good approximation in high energy hadronic collisions) the elastic dipole-target 
cross section and therefore the diffractive DIS cross section is half of the total.
In nuclei the interaction of the dipole is closer to the unitarity limit
than in protons at the same energy, and 
the fraction of diffractive events is even larger. According to 
one recent estimate~\cite{Kowalski:2008sa} 20-25\% of the events at an EIC could 
be diffractive. This is due to the combination of shadowing of the total cross section
and enhancement of the diffractive one (see \fig\ref{fig:diff}).
 There are several fascinating features in diffractive DIS. One is that such a large
fraction of the interactions in chromodynamics, even at high energy and $Q^2$ 
where asymptotic freedom should apply, happens without any net color being exchanged.
Another is the large probability of the target staying intact. In the 
target rest frame one is (at HERA energies) hitting a proton with a TeV scale electron 
without breaking it. When the same electron beam is scattered on a  nucleus the probability
of the nucleus staying intact is even larger, in spite of the huge amount of energy 
deposited compared to the typical nuclear binding energies in the $10\mev$ range.

The distribution in $t$ (the momentum kick given to the target proton or nucleus)
in diffractive DIS is directly a Fourier transform of the impact parameter distribution
of the gluons. In the proton this has led to the  observation that the gluons are 
localized in a smaller radius around the center of the proton than the valence quarks.
In the context of heavy ion collisions Monte Carlo Glauber modeling has been remarkably successful
in describing most of the transverse geometrical features of the collision system at RHIC.
There are some signals, such as $v_2$ fluctuations~\cite{Sorensen:2008zk,Alver:2008hu},
that for accurate enough
measurements one will need to understand the geometry of the actual 
small $x$ gluonic degrees of freedom better than by Glauber modeling 
which is essentially extrapolating from the valence region, since it is
based on density profiles measured from electric charge densities, and based
on an assumption of independent nucleon scatterings. Measuring diffractive observables
at an EIC will be challenging because of the smallness of the $t$ that dominate for 
large nuclei, but could significantly improve our understanding of the transverse
geometry of the small $x$ glue in nuclei.

\section{Longitudinal direction}\label{sec:long}

The ``Ridge'' structure of two particle 
correlations in central heavy ion 
collisions~\cite{Putschke:2007mi,Daugherity:2008su,Alver:2008gk,Nagle:2009wr}
and observations of long range rapidity correlations in 
particle multiplicities~\cite{Abelev:2009dq,Tarnowsky:2008am}
are striking signals of new dynamical effects at RHIC.
Because particles that are produced far away in rapidity can, by causality,
only be correlated at early times\footnote{Although when making precise estimates
one must be more careful and distinguish spacetime, momentum space and pseudorapidity
and understand shorter range correlations such as those from hadronization}, 
these correlations must have originated in the initial stages of the collision.
They should therefore be present already in the wavefunctions of the 
initial nuclei~\cite{Gelis:2008sz},
and correspond to a clearly and exactly measurable correlation observable 
in nuclear DIS. Formulating concisely and precisely what this observable would 
be is still to be done.

Another intriguing experimental observation, although still preliminary,
is event-by-event CP violation in heavy ion collisions (also discussed  at this 
conference~\cite{Voloshin:2009hr}). In the CGC framework this is naturally 
understood in terms of the same parametrically strong longitudinal 
glasma~\cite{Kharzeev:2001ev,Lappi:2006fp} fields as the long range
rapidity correlations in the ridge. The existence of the phenomenon was 
predicted~\cite{Kharzeev:2004ey} before its observation, and a more detailed
description of the mechanism is provided by the 
``Chiral Magnetic Effect''~\cite{Kharzeev:2007jp}.

\section{Energy loss in cold nuclear matter}\label{sec:cnm}

In contrast to the situation at small $x$ outlined in \se\ref{sec:sat} and discussed
in most of this talk we should also mention the opposite limit, which is relevant for
heavy ion collisions in another way. When the momentum fraction $x$ is large, the 
wavelength of the virtual photon in the target rest frame is very short.
It will not have time to fluctuate, but will instead interact \emph{locally} with 
a  quark (typically valence, since we are at large $x$). In a large nucleus, this quark 
jet will then have to travel a long way in cold nuclear matter, where it will interact 
and lose energy; see \fig\ref{fig:eloss} for one calculation compared to HERMES 
data at a much lower energy than would be reached at the EIC.
Because of the well constrained kinematics in DIS, the initial momentum of the jet is 
very precisely. Measuring the hadronization products in the process will yield 
detailed information on the process of energy loss and hadronization in cold nuclear
matter, and serve as a clear baseline and comparison for
understanding energy loss in the quark-gluon plasma. We refer to the recent 
review~\cite{Accardi:2009qv} for further details.

\section{Conclusions}\label{sec:conc}

In this talk, we have reviewed a series of topics common to the physics that could be 
studied in future DIS experiments on nuclei and the initial stage of ultrarelativistic
heavy ion collisions. We argued that in both cases, at high energy, the bulk of the 
physics is dominated by a single transverse momentum scale, the saturation scale $\qs$.
The saturation scale can be understood in different ways depending on the Lorentz frame
in which one views the process. 
On one hand it is a transverse length scale at which QCD cross
sections deviate from their perturbative rise to comply with unitarity. On the other hand
it is the transverse momentum scale at which the occupation numbers of gluonic states 
grows so large that their nonlinear interactions start to limit further radiation.
We then discussed some aspects of the high energy wave function that could be understood
at an electron ion collider with a higher precision than can be achieved by looking
at AA collisions alone. These include the numerical values for parameters, such as $\qs$,
that characterize the wavefunction (\se\ref{sec:numbers}), the transverse geometry
of the small $x$ glue in the nucleus(\se\ref{sec:tr}), the 
longitudinal  structure and correlations in rapidity present in the wavefunction 
and the correspondint structure of the longitudinal glasma fields in AA collisions
(\se\ref{sec:long}) and, finally, the local properties of the nuclear medium
as probed by a high energy jet propagating through it (\se\ref{sec:cnm}). 
This has hopefully conveyed the picture that DIS and AA experiments should
not be seen as separate domains, but complementary methods of addressing common
questions about the nature of QCD.

\section*{Acknowledgments} 
The author thanks F. Gelis, C. Gombeaud, J.-Y. Ollitrault
and R. Venugopalan for comments and discussions in preparing this talk.
The author is supported by the Academy of Finland, contract 126604.

\bibliographystyle{h-physrev4mod2}
\bibliography{spires}

\end{document}